\documentclass[prx, twocolumn, superscriptaddress,amsmath,amssymb]{revtex4-1}
\usepackage{graphicx}
\usepackage{epstopdf}
\usepackage{amsmath}
\usepackage{amssymb}
\usepackage{bm}
\usepackage{color}
\usepackage{hyperref}
\usepackage{tabularx}
\usepackage{multirow}
\usepackage{braket}
\usepackage{xcolor}
\usepackage{physics}
\usepackage{soul}
\usepackage{bbold}

\newcommand{\e}{\text{e}}

\begin{document}
\bibliographystyle{apsrev}
\title{Topological Superconductivity in Multifold Fermion Metals}
\author{Jason Z. S. Gao} \thanks{These authors contributed equally to this work}
\author{Xue-Jian Gao} \thanks{These authors contributed equally to this work}
\affiliation{Department of Physics, Hong Kong University of Science and Technology, Clear Water Bay, Hong Kong, China}
\author{Wen-Yu He}
\affiliation{Department of Physics, Massachusetts Institute of Technology, Cambridge, Massachusetts 02139, USA}
\author{Xiao Yan Xu}
\affiliation{Department of Physics, University of California at San Diego, La Jolla, California 92093, USA}
\author{T. K. Ng}
\affiliation{Department of Physics, Hong Kong University of Science and Technology, Clear Water Bay, Hong Kong, China}
\author{K. T. Law} \thanks{Corresponding author.\\phlaw@ust.hk}
\affiliation{Department of Physics, Hong Kong University of Science and Technology, Clear Water Bay, Hong Kong, China}
\date{\today}

\begin{abstract}
{Recently, multifold fermions characterized by band crossings with multifold degeneracy and Fermi surfaces with nontrivial Chern numbers have been discovered experimentally in AlPt \cite{Yulin} and XSi(X=Rh,Co) \cite{Hassan3,HongDing,Takafumi}. In this work, we largely expand the family of multifold fermion materials by pointing out that several well-studied noncentrosymmetric superconductors are indeed multifold fermion metals. Importantly, their normal state topological properties, which have been ignored in previous studies, play an important role in the superconducting properties. Taking  Li$_2$Pd$_3$B and Li$_2$Pt$_3$B as examples, we found a large number of unconventional degenerate points, such as double spin-1, spin-3/2, Weyl and double Weyl topological band crossing points near the Fermi energy, which result in finite Chern numbers on Fermi surfaces. Long Fermi arc states in Li$_2$Pd$_3$B, originating from the nontrivial band topology were found. Importantly, it has been shown experimentally that Li$_2$Pd$_3$B and Li$_2$Pt$_3$B are fully gapped and gapless superconductors, respectively. By analyzing the possible pairing symmetries, we suggest that  Li$_2$Pd$_3$B can be a DIII class topological superconductor with Majorana surface states, even though the spin-orbit coupling in  Li$_2$Pd$_3$B is negligible.  Interestingly, Li$_2$Pt$_3$B, being gapless, is likely to be a nodal topological superconductor with dispersionless surface Majorana modes. We further identified that several noncentrosymmetric superconductors, such as Mo$_3$Al$_2$C, PdBiSe, Y$_2$C$_3$ and La$_2$C$_3$, are multifold fermion superconductors whose normal state topological properties have been ignored in previous experimental and theoretical studies.} 
\end{abstract}

\maketitle
\section{Introduction}
Noncentrosymmetric superconductors are superconductors without inversion centers. The broken inversion symmetry results in antisymmetric spin-orbit coupling (ASOC) which can lead to interesting superconducting properties, such as mixed pairing order parameters \cite{Gorkov1,Sigrist1,Sigrist3}, helical phases \cite{Agterberg2003,Sigrist2005,Feigelman2003}, novel magnetoelectric responses \cite{Mukuda,Yogi,Kimura,Lu2015}, large enhancement of the upper critical field $H_{c2}$ \cite{Lu2015,Xi2016} or topological superconductivity \cite{Nagaosa2012,Schmalian2015,Samokhin2015,Zhang2011,Fujimoto2010}. In the past two decades, many noncentrosymmetric superconductors have been discovered \cite{Smidman}. To understand the superconducting properties of noncentrosymmetric superconductors, a common procedure was to construct normal state Hamiltonians which respect the crystal symmetry, and then to further include ASOC and study its effect on superconductivity. However, the normal state topological properties of these noncentrosymmetric superconductors have usually been ignored in previous studies.

In recent years, tremendous progress had been made in the understanding of the topological properties of band structures. Particularly, many topologically nontrivial band crossings which describe unconventional fermions, such as Weyl, Kramers Weyl and unconventional multifold fermions have been discovered \cite{Bernevig1,Hassan1,Shoucheng,Hassan2,Hassan3,HongDing,Yulin,Takafumi}. These multifold band crossings give rise to finite Chern numbers on Fermi surfaces enclosing the band crossing points. The case of multifold fermions is particularly interesting, as the nonsymmorphic and time-reversal (TR) symmetry can enforce multiple degeneracies and result in large Chern numbers on Fermi surfaces enclosing the band crossing points, even in the absence of ASOC. Unlike Weyl points, which usually result in short Fermi arc states, multifold fermion crossing points result in incredibly long Fermi arc states which span a large portion of the surface Brillouin zone. So far, two multifold fermion semimetals, namely, AlPt \cite{Yulin} and XSi(X=Rh,Co) \cite{Hassan3,HongDing,Takafumi}, have been identified, and their long Fermi arc states have been observed through ARPES experiments recently. However, they are not superconducting, and it is not known how the normal state topology resulting from multifold fermions can affect the superconducting properties in realistic materials. 

In this work, we point out that Li$_2$Pd$_3$B and Li$_2$Pt$_3$B, which are well-studied noncentrosymmetric superconductors \cite{Hirata,Yuan,ZhengGuoqing1,ZhengGuoqing2}, are superconducting multifold fermion metals and the normal state topological properties of these materials has been ignored in previous studies. Through NMR \cite{ZhengGuoqing1,ZhengGuoqing2,ZhengGuoqing3}, specific heat \cite{SpecificHeat,SpecificHeat2} and penetration length \cite{Yuan} measurements, it has been suggested that Li$_2$Pd$_3$B and Li$_2$Pt$_3$B are fully gapped and gapless superconductors, respectively.  In the case of Li$_2$Pd$_3$B, there exist 6-fold crossings at the R-point of the Brillouin zone, and there are isolated hole pockets enclosing the R-points at the Fermi energy \cite{ZhengGuoqing2,Pickett}. The R-pocket carries a Chern number of 4 \cite{Bernevig1}, and the rest of the Fermi surfaces carry a total Chern number of $-4$. When the pairing of the R-pocket and the rest of the Fermi surfaces have opposite pairing signs, the material can be a fully gapped DIII class topological superconductor with four Majorana cones on the (001) surfaces. It is important to note that the finite Chern number of the R-pocket is generated by the multifold fermion which does not originate from ASOC. This is very different from other noncentrosymmetric superconductors in which the unconventional and topological properties originate from ASOC \cite{Nagaosa2012,Schmalian2015,Fujimoto2010,Schnyder2,Ryu,Schnyder2012}. Moreover, we found long Fermi arcs on the (001) surfaces which span long distances in the surface Brillouin zone. 

In the case of Li$_2$Pt$_3$B, the superconducting state is gapless. The order parameters belonging to both the A$_1$ (the isotropic representation) and A$_2$ representations of the O point group can result in gapless superconducting phases. Importantly, the superconducting states can be topological and possess dispersionless Majorana modes on surfaces in both A$_1$ and A$_2$ representations. Therefore, we suggest that Li$_2$Pt$_3$B is a promising gapless topological superconductor candidate. 

In this manuscript, we first identify topological band crossing points in Li$_2$Pd$_3$B and Li$_2$Pt$_3$B which are relevant to the Fermi surface topology through \textit{ab initio} calculations. Several 2-fold, 4-fold and 6-fold band crossings which correspond to Weyl, Kramers Weyl, spin-1 and multifold fermions were found and are listed in Table~\ref{ChiralFermions}. Second, by introducing extended s-wave pairing to the bands near the Fermi energy, we show that a fully gapped DIII topological superconducting phase in Li$_2$Pd$_3$B can be obtained. Four Majorana cones on the (001) surface were found. Third, we show that the gapless superconducting states of the A$_1$ and A$_2$ representations, which are relevant to Li$_2$Pt$_3$B, are topological and possess dispersionless Majorana surface modes.  Finally, we point out that a few other well studied noncentrosymmetric superconductors, such as Mo$_3$Al$_2$C , PdBiSe, Y$_2$C$_3$ and La$_2$C$_3$ are also multifold fermion metals with rich topological band structures. These materials provide new platforms to study the interplay between normal state topology and superconductivity.  

		\begin{figure}
		  \centering
		      \includegraphics[width=0.45\textwidth]{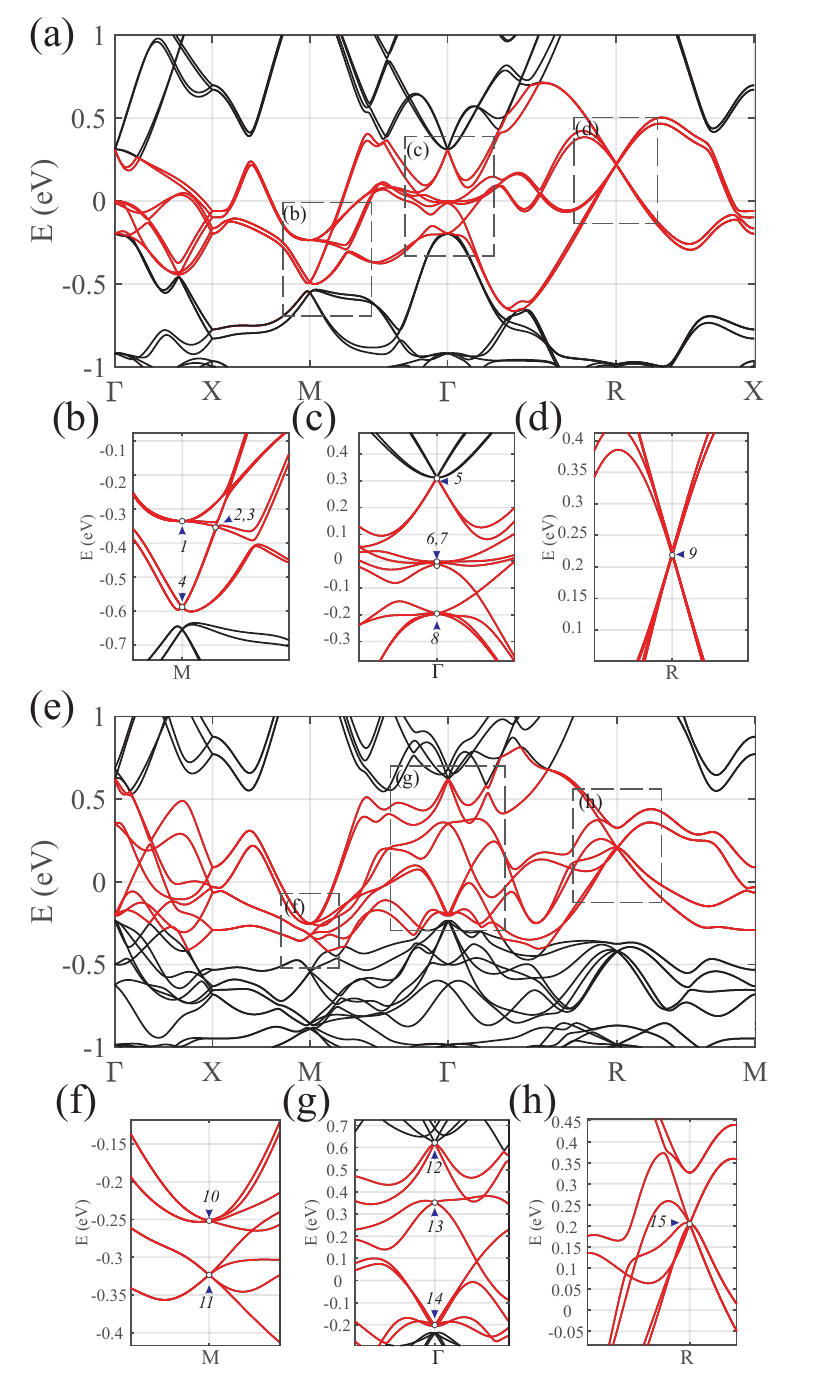}
		   \caption{{\bf(a-d)} Band structure of Li$_2$Pd$_3$B. {\bf(e-h)} Band structure of the Li$_2$Pt$_3$B. Bands which cross the Fermi energy are marked in red, and the topological band crossing points which can be described by unconventional fermions are depicted in enlarged plots. The degeneracies corresponding to unconventional fermions are enumerated and listed in Table~\ref{ChiralFermions}.}\label{Fig1}
		\end{figure}

		\begin{figure}
		  \centering
		      \includegraphics[width=0.45\textwidth]{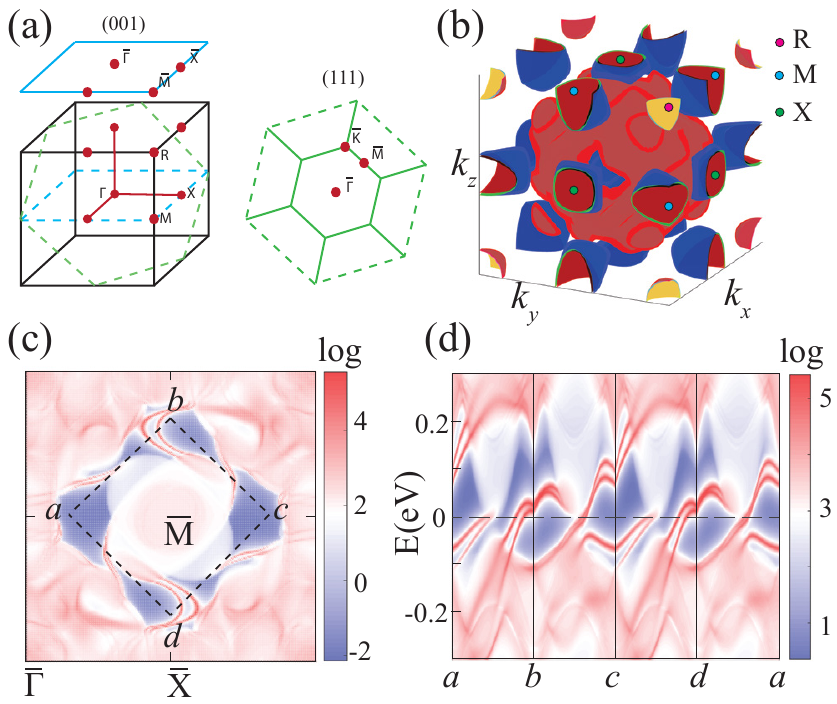}
		   \caption{{\bf(a)} The bulk, the $(001)$ surface and the $(111)$ surface Brillouin zones of a cubic lattice. {\bf(b)} The Fermi surface of Li$_2$Pd$_3$B. It can be seen that there are Fermi pockets enclosing all of the TRIMs.  {\bf(c)} Surface Fermi arcs of Li$_2$Pd$_3$B. The colormap represents the logarithm of spectral weight on the (001) surface. Near the Fermi energy, there are 12 Fermi arcs near the $\mathrm{\bar{M}}$ pocket. {\bf(d)} The energy dependence of the $k$-points of a closed loop is indicated by dashed lines in {\bf(c)}. The energy spectral flow is calculated, revealing that the total topological charge of $C = 12$.}\label{Fig2}
		\end{figure}

\section{Multifold Fermions in $\text{Li}_2\text{Pd}_3\text{B}$ and $\text{Li}_2\text{Pt}_3\text{B}$} 
Both Li$_2$Pd$_3$B and Li$_2$Pt$_3$B have a perovskite-like cubic structure, and they belong to the nonsymmorphic space group P4$_3$32 (No. 212) with point group O \cite{Pickett}. In the reciprocal space, the cubic Brillouin zone has nonequivalent high symmetry points $\Gamma\left(0, 0, 0\right)$, X$\left(\pi/a, 0, 0\right)$, M$\left(\pi/a, \pi/a, 0\right)$ and R$\left(\pi/a, \pi/a, \pi/a\right)$ (with primitive lattice constant $a$), as shown in Fig.~\ref{Fig2}a. {\it Ab initio} calculations (Appendix~\ref{APa}) are used to obtain the band structures of Li$_2$Pd$_3$B and Li$_2$Pt$_3$B, as depicted in Fig.~\ref{Fig1}. The band structures of Li$_2$Pd$_3$B and Li$_2$Pt$_3$B were calculated more than a decade ago \cite{Pickett}, but the unconventional crossing points that correspond to the multifold fermions have not been identified until recently. It was pointed out in \cite{Bernevig1} that Li$_2$Pd$_3$B possesses a six-fold degenerate point at the R point even though the interplay between superconductivity and normal state topology was not studied. 

To be more specific, the generators of the space group P4$_3$32 (in Seitz symbols) are
\begin{equation}
	\begin{aligned}
		&\qty{2_{001}|\tfrac{1}{2},0,\tfrac{1}{2}},   &\qquad\qty{2_{010}|0,\tfrac{1}{2},\tfrac{1}{2}}, \\
		&\qty{3^+_{111}|0}, &\qquad\qty{2_{110}|\tfrac{1}{4},\tfrac{3}{4},\tfrac{1}{4}}.
	\end{aligned}
\end{equation}
These crystal symmetries, together with time-reversal symmetry $\mathcal{T}$, will stabilize a 6-fold degeneracy at $\Gamma$ and 8-fold degeneracy at $R$ in the absence of SOC. In the presence of SOC, the 6-fold symmetry will split into 4-fold and 2-fold degenerate points, while the 8-fold degeneracy at $R$ will split into 2-fold and 6-fold degenerate points. The effects of SOC can be seen by comparing the band structures in Fig.~\ref{Fig1}, as Li$_2$Pd$_3$B has much weaker SOC than Li$_2$Pt$_3$B. It is important to note that the 6-fold degenerate point at R is topologically nontrivial \cite{Bernevig1}. It is described by the two spin-1 fermions, each of which gives rise to quantized Berry flux of $\pm 2$ for a gapped surface enclosing R.
Here the spin-$n$ fermions are defined to represent the electronic states around a specific $k$-point that transform under the crystalline rotational symmetries in the form of $\exp(-i\hat{\bm{S}}\cdot\hat{\bm{n}}\theta)$, where $\hat{\bm{S}}$ is the spin-$n$ angular momentum operator and the unit vector $\hat{\bm{n}}$ and $\theta$ represent the axis and angle of a rotation operation respectively.
In Li$_2$Pd$_3$B and Li$_2$Pt$_3$B, time-reversal symmetry forces two copies of spin-1 fermions to be degenerate at $R$ \cite{Bradley}, giving a 6-fold degeneracy as depicted in Fig.~\ref{Fig1}d. This is referred to as double spin-1 in Table~\ref{ChiralFermions}. On the other hand, the 4-fold degenerate fermions at $\Gamma$ (e.g. labeled as No.~14 in Fig.~\ref{Fig1}g) are the spin-3/2 or the so-called Rarita-Schwinger-Weyl (RSW) fermions with a monopole charge of $\pm4$. The corresponding irreducible representations and an effective $k\cdot p$ Hamiltonian linear in $k$ for the two aforementioned unconventional fermions are included in Appendix~\ref{APe}. 

Kramers Weyl fermions at $M$, which is a TRIM, are doubled by nonsymmorphic symmetries, giving double Kramers Weyl fermions \cite{Bradley}. Two Weyl fermions (No.~2,3 in Table~\ref{ChiralFermions}) along the M-$\Gamma$ line with a total Berry curvature charge of $+2$ have to be taken into account for the observed surface spectrum. For the bands which cross the Fermi energy (denoted in red in Fig.~\ref{Fig1}), 15 unconventional degenerate points are identified as unconventional chiral fermions. The locations of these points are labeled in Fig.~\ref{Fig1}, and the degeneracy and the topological charges of the Fermi surfaces enclosing these points are summarized in Table~\ref{ChiralFermions}.

\begin{table}
	\caption{Unconventional fermions found in Li$_2$Pd$_3$B and Li$_2$Pt$_3$B by first-principles calculation. Indices correspond to those in Fig.\ref{Fig1}.} 
	\centering 
	\begin{tabular}{p{3.5cm} p{1.5cm} p{1.7cm} p{1.5cm} } 
	\hline\hline 
	Type & Monopole charge & Degeneracy& No.\\ [0.5ex] 
	\hline 
	Weyl& $\pm1$ & 2 & 2,3 \\
	Kramers Weyl & $\pm1$ & 2 & 6,7,13 \\
	Double Kramers Weyl & $\pm2$ & 4 & 1,4,10,11 \\
	Double Spin-1 & $\pm4$ & 6 & 9,15 \\
	Spin-3/2 & $\pm4$ & 4 & 5,8,12,14 \\[1ex]
	\hline 
	\end{tabular}
	\label{ChiralFermions} 
\end{table}

When a nondegenerate Fermi surface in the Brillouin zone encloses the band crossing points, it can acquire a finite Chern number, and there can be surface Fermi arc states connecting surfaces with different Chern numbers. The Fermi surface of Li$_2$Pd$_3$B is depicted in Fig.~\ref{Fig2}b. Fig.~\ref{Fig2}c visualizes the calculated surface spectral function 
$\mathcal{A}_s(E,\bm{k})=-\frac{1}{\pi}\Im G_s(E,\bm{k})$
of the (001) surface of Li$_2$Pd$_3$B, where $G_s(E,\bm{k})$ is the surface Green's function. As many as 12 Fermi arcs emerge from the $\bar{M}$ pocket. The $\bar{M}$ surface pocket is a projection of the $R$ and $M$ pockets of the 3D Brillouin zone, and the rest of the states enclosing the $\bar{M}$ pocket are projections of pockets around the $\Gamma$ point in the 3D Brillouzin zone. The energy dispersion of the Fermi arc states are depicted in Fig.~\ref{Fig2}d. Note that all the Fermi arc states around the $\bar{M}$ pocket have positive Fermi velocity along the a-b-c-d lines defined in Fig.~\ref{Fig2}c. This suggests that the gapped surface enclosing the M and R points in the 3D Brillouin zone has a total Chern number of 12. This is consistent with the fact that the two Fermi surfaces at the R point carry a total monopole charge of 4 and the eight Weyl points (No.~2,3 in Table~\ref{ChiralFermions} together with Weyl points related by 4-fold rotation symmetry) near M carry a total monopole charge of 8. 

\section{Pairing in superconducting $\text{Li}_2\text{(Pd}_x\text{Pt}_{1-x}\text{)B}$}

Previous experiments have shown that Li$_2$Pd$_3$B and Li$_2$Pt$_3$B are superconducting with T$_c$ at $7.5K$ and $2.13K$, respectively \cite{Yuan, ZhengGuoqing1, ZhengGuoqing2}. Through NMR \cite{ZhengGuoqing1,ZhengGuoqing2,ZhengGuoqing3}, specific heat and penetration depth measurements \cite{SpecificHeat,SpecificHeat2,Yuan}, Li$_2$Pd$_3$B was found to be a fully gapped superconductor. When Pd atoms are gradually replaced by Pt atoms, the pairing gap narrows, and the material becomes gapless when $x < 0.2$ in Li$_2$(Pd$_x$Pt$_{1-x}$)$_3$B and remains gapless for Li$_2$Pt$_3$B \cite{ZhengGuoqing2,ZhengGuoqing3}. With these exotic properties, Li$_2$(Pd$_x$Pt$_{1-x}$)$_3$B are promising platforms for studying the interplay between superconductivity and the topological properties of multifold fermions. However, the pairing symmetry form and the topological properties of their superconducting phases are yet to be ascertained.

In the following, we analyze the possible pairing symmetries of Li$_2$(Pd$_x$Pt$_{1-x}$)$_3$B from their space group P$4_3$32. As inversion symmetry breaking lifts the band degeneracies at general $\bm{k}$ points in the Brillouin zone, except at TRIMs and Brillouin zone boundaries (due to nonsymmorphic symmetry), stable pairing on the Fermi surface is formed by the Bloch states $\ket{\phi_{\nu, \bm{k}}}$ and their time-reversal partners $\ket{\tilde{\phi}_{\nu, \bm{k}}}=\mathcal{T}\ket{\phi_{\nu, \bm{k}}}=e^{i\theta(\bm{k})}\ket{\phi_{\nu, -\bm{k}}}$ with a gauge-dependent phase $\theta(\bm{k})$. The pairing order parameter can be written as $\Delta_{\nu, \bm{k}}=\mel{\phi_{\nu, \bm{k}}}{\hat{\Delta}}{\tilde{\phi}_{\nu, \bm{k}}}$ \cite{Sigrist2} as the projected pairing potential $\hat{\Delta}$ on the $\nu$-th band. As a result, the Bogliubov-de Gennes Hamiltonian in the band basis takes the form (Appendix~\ref{APc})
\begin{align}
\mathcal{H}_{\textrm{BdG}}\left(\bm{k}\right)=\sum_{\bm{k}, \nu}\begin{pmatrix}
\phi^\dagger_{\nu, \bm{k}} & \tilde{\phi}_{\nu, \bm{k}}
\end{pmatrix}\begin{pmatrix}
\xi_{\nu, \bm{k}} & \Delta_{\nu, \bm{k}} \\
\Delta^*_{\nu, \bm{k}} & -\xi_{\nu, \bm{k}}
\end{pmatrix}\begin{pmatrix}
\phi_{\nu, \bm{k}} \\ \tilde{\phi}^{\dagger}_{\nu, \bm{k}}
\end{pmatrix}.
\label{BdG}
\end{align}
With the crystalline symmetry of Li$_2$Pd$_3$B and Li$_2$Pt$_3$B, the pairing $\Delta_{\nu, \bm{k}}$ transforms as one of the irreducible representations (IR) of the point group O, which is the factor group of P$4_3$32. This gives the symmetry of the pairing function, which decides the gap structure of quasi-particle spectrum \cite{RevModPhy,volovik,Venderbos,Brydon}. 
Using this method, we can classify all the symmetry-allowed pairing functions up to their leading order in $k$ corresponding to the irreducible representations of point group O, as listed in the Table~\ref{tb3} of Appendix~\ref{APc}.

We found that the space group P$4_3$32 allows two one-dimensional representations, $A_1$ and $A_2$, as listed in Table~\ref{ChiralFermions}. The $A_1$ representation includes the conventional s-wave and the extended s-wave pairing. As we will show below, the extended s-wave pairing can result in a fully gapped DIII class topological superconducting phase with four Majorana cones on a (001) surface. Therefore,  Li$_2$Pd$_3$B can either be a fully gapped topologically trivial superconductor or a DIII class topological superconductor. For Li$_2$Pt$_3$B, with its gapless pairing, the material cannot be a fully gapped s-wave superconductor. To obtain the gapless energy spectrum, Li$_2$Pt$_3$B needs to be either in the $A_2$ pairing phase or in the gapless regime of the $A_1$ pairing phase. Importantly, both of these possible gapless pairings lead to topological phases and the surface Majorana modes. Therefore, Li$_2$Pt$_3$B is a promising gapless topological superconductor candidate. Space group P4$_3$32 also allows higher dimensional representations, as discussed in the Appendix~\ref{APc}. Nevertheless, the pairings belonging to the higher dimensional representations either break crystal symmetries or time-reversal symmetry, and the breaking of these symmetries has not been observed experimentally. Therefore, in this work, we focus on pairings associated with the one-dimensional representations.

	\begin{table}
		\caption{Possible pairing phases of Li$_2$Pt$_3$B labeled by the 1D IR of O point group. In each IR, the pairing function $\Delta_{\nu, \bm{k}}$ can be expanded in terms of the basis functions. The $\eta_\zeta$ with $\zeta=1, 2, 3, ...$ there are the coefficients for the basis functions in A$_1$ and A$_2$ IR respectively.} 
		\centering 
		\begin{tabularx}{0.485\textwidth}{c c c c c c} 
		\hline\hline 
		IR & Pairing function $\Delta_{\nu, \bm{k}}$ \\ [0.5ex] 
		\hline 
		A$_1$ & $\eta_1$+$\eta_2\left(3-\cos k_xa-\cos k_ya-\cos k_za\right)+...$ \\
		A$_2$ &{\footnotesize $\eta_3 (\cos k_xa-\cos k_ya)(\cos k_ya-\cos k_za)(\cos k_za-\cos k_xa)+...$}\\[1ex] 
		\hline 
		\end{tabularx}
		\label{Pairing_IR} 
	\end{table}

	\begin{figure}
		\centering
		\includegraphics[width=0.45\textwidth]{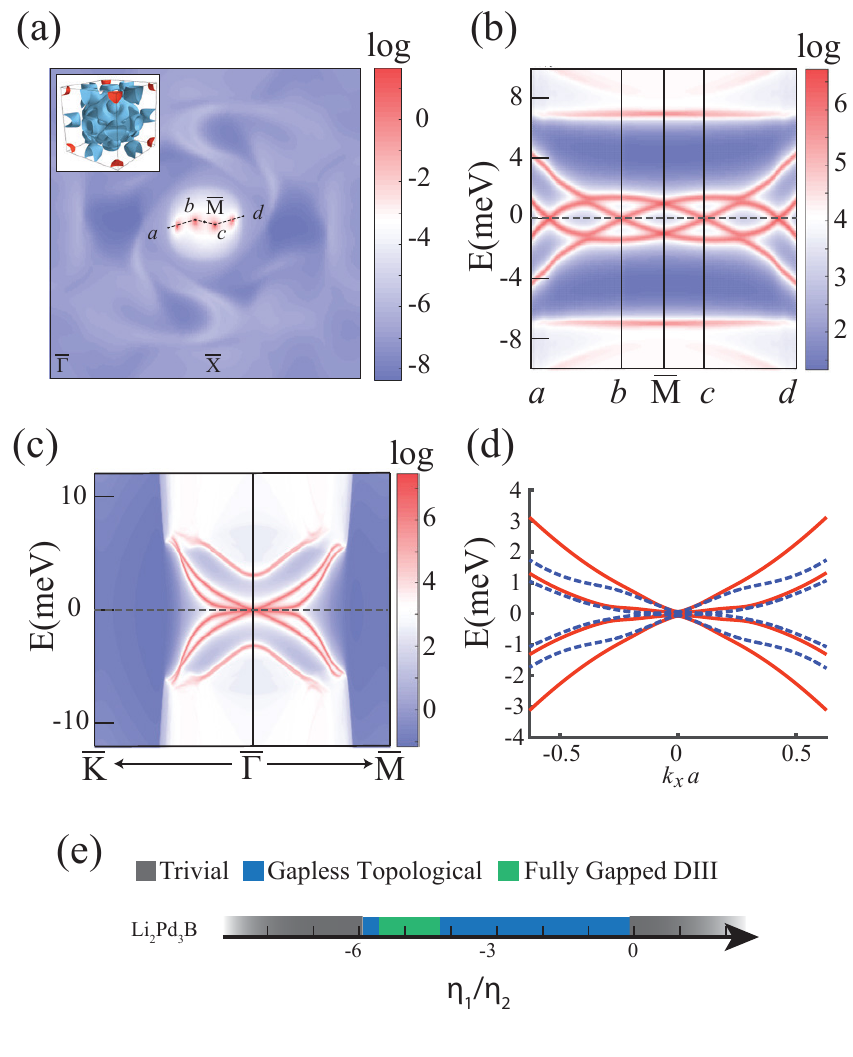}
		\caption{{\bf(a)} Four Majorana cones on the (001) surface of Li$_2$Pd$_3$B in the DIII topological phases with fully gapped $A_1$ pairing. The colormap represents the logarithm of quasi-particle spectral function weight on the (001) surface at $E=0$. 
		The inset indicates that the sign of projected pairing potential is positive (red) on the $R$ pocket and negative (blue) on the rest. This pairing corresponds to $\eta_1/\eta_2=-5,\eta_2=10$ meV, leading to an $N_{\text{gap}} = 4$ phase. 
		{\bf(b)} A zig-zag cut passing through all the Majorana cones is made and the corresponding energy spectrum is calculated.
		{\bf(c)} and {\bf(d)} show the energy spectrum of Majorana states on the (111) surface of  Li$_2$Pd$_3$B in its DIII topological superconducting phase. 
		The red solid lines and blue dashed lines in {\bf(d)} represent the respective surface Majorana bands on the two sides. The different dispersion on the two sides is due to the breaking of inversion. Clearly on both (111) surface sides, there are two Majorana cones with linear and cubic dispersions.
		{\bf(e)} Different $\eta_1/\eta_2$ ratios in the A$_1$ pairing shown in Table~\ref{Pairing_IR} give different topological superconducting phases.
		}\label{Fig3}
	\end{figure}

\section{Fully gapped A\textsubscript{1} pairing phase in $\text{Li}_2\text{Pd}_3\text{B}$} 
As shown in the previous sections, the band structure of Li$_2$Pd$_3$B is highly topologically nontrivial, and the Fermi surface topological properties cannot be captured by simple effective models. Therefore, in our calculations, we first construct the Wannier orbitals from first-principles calculations with the \textit{Wannier90} package \cite{Mostofi2014} and the Hamiltonian is diagonalized to obtain the eigenstates  $\ket{\phi_{\nu, \bm{k}}}$. This allows us to incorporate the topological properties of the normal state Fermi surface when studying the superconducting phase. For simplicity, we assign the same $\Delta_{\nu, \bm{k}}$ to all the bands and obtain the Bogliubov-de Gennes Hamiltonian in the form of Eq.~\ref{BdG}. For the case of Li$_2$Pd$_3$B, with the A$_1$ representation, the superconducting properties of the system depend on $\eta_1$ and $\eta_2$ of the order parameter $\Delta_{\nu, \bm{k}} = \eta_1+\eta_2\left( 3 - \cos k_xa - \cos k_ya - \cos k_za \right)$. When $|\eta_1/\eta_2|\gg1$, the pairing has the same sign in the whole Brillouin zone and the material is a topologically trivial superconductor. Interestingly, there is a large parameter regime, as shown  Fig.~\ref{Fig3}e, where the pairing of the large Fermi pockets near $\Gamma$ and the two Fermi pockets enclosing the R point have different signs. In this case, a DIII class topological superconductor is obtained \cite{Ludwig2008}. The topological invariant of the DIII class topological superconductor is given by
	\begin{equation}\label{DIII}
		N_{\textrm{DIII}}=\frac{1}{2}\sum_\nu\textrm{sgn}\left(\Delta_{\nu, \bm{k}_{\textrm{F}}}\right)C_\nu,
	\end{equation}
where $\Delta_{\nu, \bm{k}_{\textrm{F}}}$ is the pairing on the Fermi surface of the $\nu$-th band with Chern number $C_\nu$ \cite{Xiaoliang}. 

In the topological regime, as the two nearly degenerate Fermi pockets enclosing the R point have a total Chern number of 4 and the same pairing sign, and the rest of the Fermi surfaces have a total Chern number of $-4$ which is opposite to the R pocket, the total topological invariant of the DIII class topological phase is $N_{\textrm{DIII}} =4$. As a result, we expect that four Majorana cones appear on the (001) surface of the material. This is verified numerically by setting $\eta_1 / \eta_2 = -5$, and four zero energy modes are found on the Brillouin zone of the (001) surface state. It is interesting to note that on the (111) surface, the $C_3$ rotational symmetry forces the Majorana cones to be pinned at the surface Brillouin zone center and result in two Majorana cones with linear and cubic dispersions, as shown in Fig.~\ref{Fig3}c and \ref{Fig3}d. This cubic dispersion Majorana cone is a manifestation of the existence of multifold fermions, particularly the spin-3/2 fermions, as pointed out in \cite{Fang2015,Yang2016}. We would like to emphasise that the topological superconducting phase in Li$_2$Pd$_3$B can be achieved with an extended s-wave pairing due to the topologically nontrivial normal state band structures. This is in sharp contrast to the DIII topological superconductors proposed in \cite{Smidman,Yang2016,Fang2015,Venderbos,Fu2010,Sato2010}, which require unconventional spin-triplet pairings.  

 \begin{figure}
  \centering
      \includegraphics[width=0.45\textwidth]{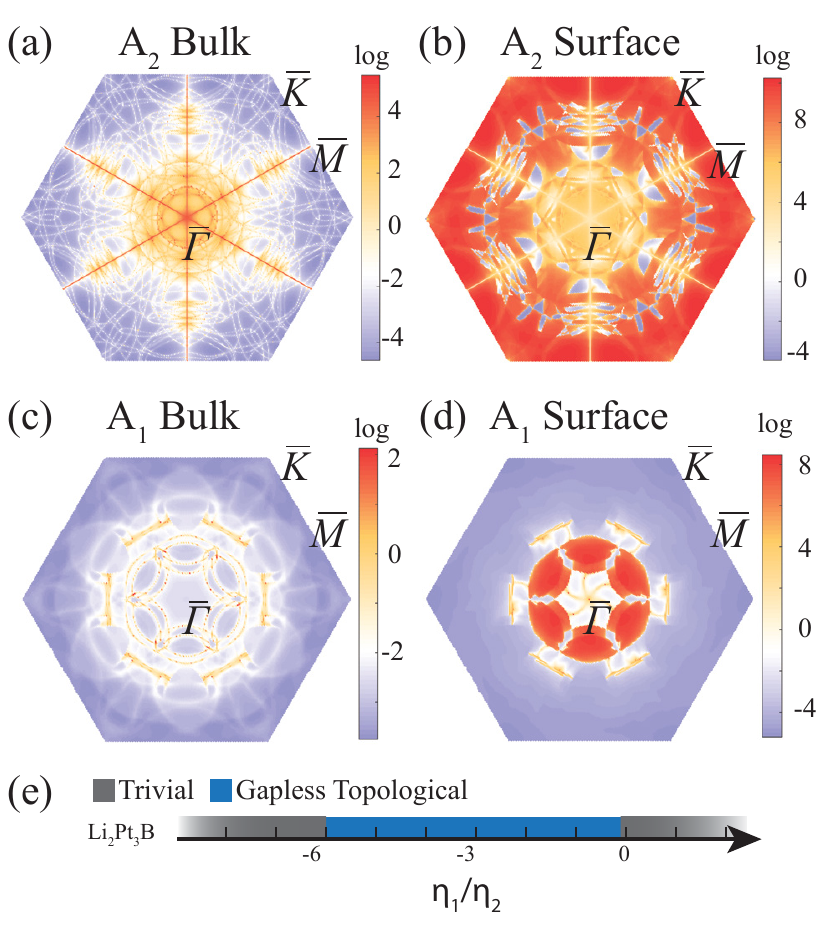}
   \caption{ Gapless topological phases of Li$_2$Pt$_3$B. {\bf(a)} The projected bulk spectral function in the $(111)$ direction at $E=0$ in the A$_2$ phase, with $\eta_3=0.16eV$ . {\bf(b)} The surface spectral function of the $(111)$ surface at $E=0$ in the A$_2$ phase.  Nodal lines shown in {\bf{(a)}} separate the surface Majorana pockets in {\bf{(b)}}. {\bf(c)} Bulk spectral functions in the A$_1$ phase. {\bf(d)}   The surface spectral function of the $(111)$ surface in the A$_1$ phase. {\bf(e)} The phase diagram for the A$_1$ phase with different $\eta_1/\eta_2$. }\label{Fig4}
\end{figure}

\section{Gapless A\textsubscript{2} pairing in $\text{Li}_2\text{Pt}_3\text{B}$}
On the other hand, Li$_2$Pt$_3$B has been experimentally found to be a gapless superconductor \cite{SpecificHeat,SpecificHeat2,Yuan}. From the symmetry analysis, we find that the $A_2$ pairing phase with $\Delta_{\nu, \bm{k}} = \eta_3 (\cos k_xa-\cos k_ya)(\cos k_ya-\cos k_za)(\cos k_za-\cos k_xa)$ satisfies the gapless condition. It is evident from the pairing function that there are many nodal planes such as the $k_x=k_z$ plane. When the nodal planes intersect with the Fermi surface, nodal lines are formed. The projection of the nodal lines on the surface Brillouin zone can create pockets which are filled by dispersionless Majorana modes on the (111) surfaces, as shown in Fig.~\ref{Fig4}. To understand the topological origin of the dispersionless Majorana modes on the (111) surface, we can parameterize the Hamiltonian as $\mathcal{H}_{\textrm{BdG}}(\bar{k}_x, \bar{k}_y, \bar{k}_z)$, where $\bar{k}_z $ is along the (111) direction and $\bar{k}_x$ and $\bar{k}_y$ are momenta parallel to the (111) surface. By taking $\bar{k}_x$ and $\bar{k}_y$ as numbers, $\mathcal{H}_{\textrm{BdG}}(\bar{k}_x, \bar{k}_y, \bar{k}_{z})$ can be regarded as a 1D Hamiltonian $\mathcal{H}_{\textrm{BdG},\bar{k}_x, \bar{k}_y}(\bar{k}_z)$ which depends on $\bar{k}_{z}$ only. The topological invariant can be written as:
\begin{equation} \label{1D}
N_{\bar{k}_x, \bar{k}_y} = \frac{1}{2\pi}\Im\int\dd{\bar{k}_z}\Tr[\partial_{\bar{k}_z}\ln Q(\bar{k}_x, \bar{k}_y, \bar{k}_z)],
\end{equation}
which determines the number of zero energy Majorana modes at the point $(\bar{k}_x, \bar{k}_y)$ of the surface Brillouin zone \cite{Schnyder2,Ryu}. Here, $Q(\bar{k}_x, \bar{k}_y, \bar{k}_z)=H_0(\bar{k}_x, \bar{k}_y, \bar{k}_z)+i\Delta(\bar{k}_x, \bar{k}_y, \bar{k}_z)$ and $H_0(\bar{k}_x, \bar{k}_y, \bar{k}_z)$ is the normal state Hamiltonian.

It is important to note that the A$_1$ pairing phase can also result in a gapless spectrum. As shown in Fig.~\ref{Fig4}c and \ref{Fig4}d, there is a wide range of parameter regime for $\eta_1/\eta_2$ where the system is gapless. Importantly, there are also dispersionless surface Majorana modes on the surfaces, as shown in Fig.~\ref{Fig4}d. The topological origin of the Majorana modes can also be determined by Eq.~\ref{1D}. Therefore, we conclude that the multifold fermion material Li$_2$Pt$_3$B is likely to be a gapless topological superconductor with Majorana surface modes.

\begin{figure}
  \centering
      \includegraphics[width=0.5\textwidth]{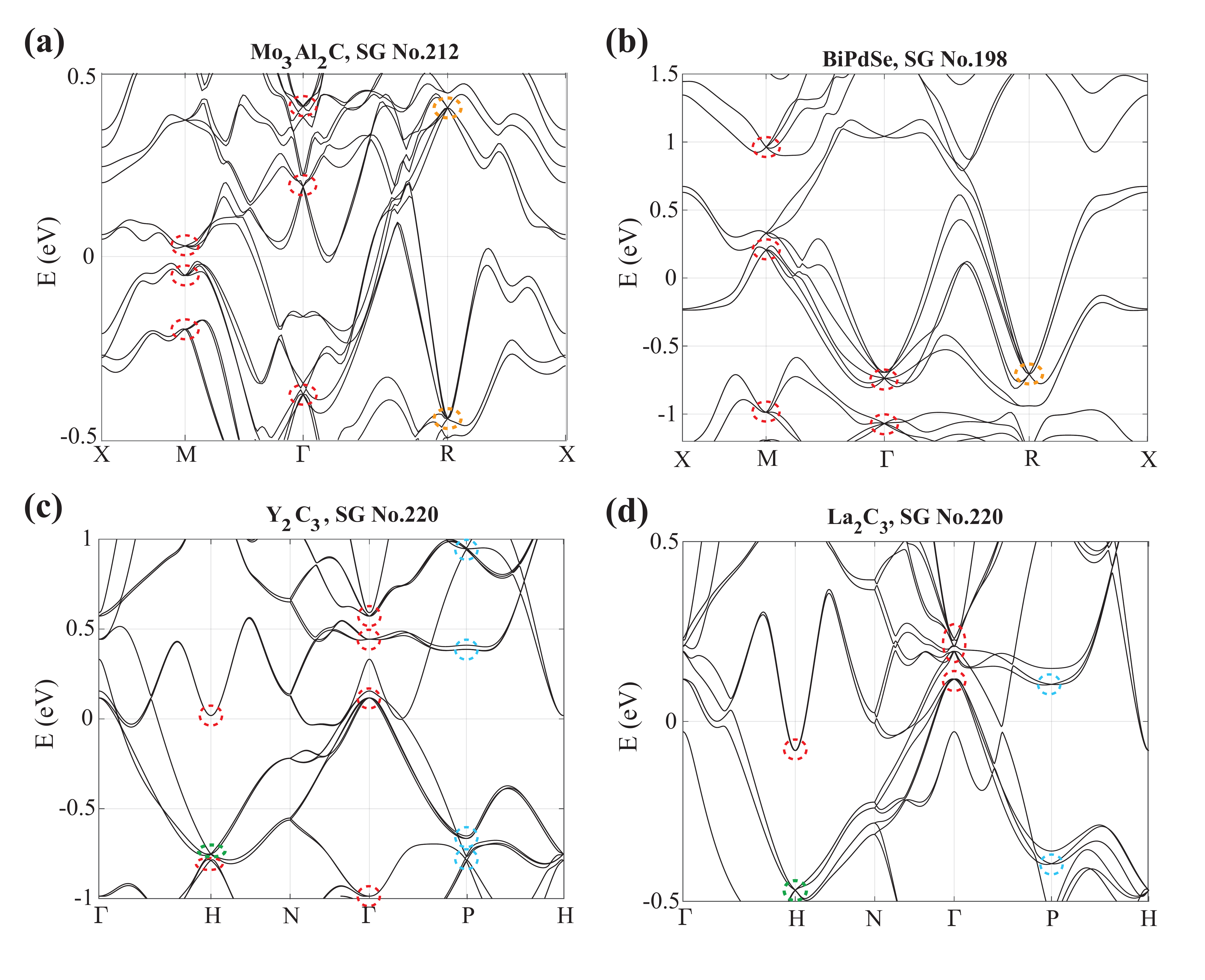}
   \caption{ The first-principles band structure of other multifold fermion superconductors, Mo$_3$Al$_2$C, BiPdSe, Y$_2$C$_3$ and La$_2$C$_3$. The blue, red, orange and green dashed circles highlight the 3-, 4-, 6- and 8-fold fermions, respectively. }\label{Fig5}
\end{figure}

\section{Other superconducting materials with multifold fermions}
In this work, we show that the normal state topology plays an important role in determining the superconducting properties of Li$_2$Pd$_3$B and Li$_2$Pt$_3$B. We expect that the normal state topological  properties are important in many other noncentrosymmetric superconductors as well. By revisiting the space group of noncentrosymmetric superconductors \cite{Smidman,Karki2010,Joshi2015,Krupka1969,Simon2004}, we can further identify several superconductors with multifold fermions, such as Mo$_3$Al$_2$C, PdBiSe, Y$_2$C$_3$ and La$_2$C$_2$ whose normal state topological properties have been ignored in previous studies.
Their first-principles band structures are shown in Fig.~\ref{Fig5}. Due to their nonsymmorphic symmetries, plentiful unconventional fermions with 3-, 4-, 6- and 8-fold degeneracy appear at high-symmetry $k$ points, and some of them are near the Fermi level and thus influence the superconducting phases. Their possible topological superconducting phases can be further studied with the same numerical and theoretical methods presented in this work.

\section{Conclusion}
In this work, we showed that Li$_2$Pd$_3$B and Li$_2$Pt$_3$B are multifold fermion metals with unconventional degenerate points at time-reversal invariant momenta due to nonsymmorphic symmetry. This results in Fermi surfaces with higher Chern numbers, even in the absence of ASOC. We showed that there are long Fermi arc states in Li$_2$Pd$_3$B at the Fermi energy. Interestingly, Li$_2$Pd$_3$B, being a fully gapped superconductor, is a candidate material of a DIII class topological superconductor. On the other hand, superconducting Li$_2$Pt$_3$B, being gapless, is likely to be a nodal topological superconductor with dispersionless Majorana surface modes. We also identified several other superconducting multifold fermion superconductors, which can also provide new platforms to study the interplay between normal state topology and superconductivity. 

\begin{acknowledgements}
The authors thank Mengli Hu,  Ruopeng Yu and Guo-qing Zheng for valuable discussions. KTL acknowledges the support of the Croucher Foundation, the Dr. Tai-chin Lo Foundation and the HKRGC through grants C6026-16W, C6025-19G, 16310219 and 16309718.
\end{acknowledgements}

\appendix

\section{First-principles calculation of band structures and surface states for Li$_2$Pd$_3$B and Li$_2$Pt$_3$B}\label{APa}

  All the density functional theory (DFT) \cite{dft1} calculations were performed with \textit{the Vienna Ab initio Simulation Package} (VASP) \cite{dft2} with the projector-augmented wave method \cite{dft3} and the exchange-correlation functional of the Perdew-Berke-Ernzerhof (PBE) form \cite{dft4} in the generalized-gradient approximation \cite{dft5}. Before we calculated the band structures, the atomic structures of Li$_2$BPd$_3$ and Li$_2$BPt$_3$ were fully relaxed with the maximum residual force $10^{-4}$ eV$\cdot$\AA$^{-1}$. The Monkhorst-Pack k-point sampling of $9\times9\times9$ was adopted for the Brillouin zone integration in the relaxation as well as for the self-consistent calculation. 
  The Wannier tight-binding Hamiltonian was obtained with the \textit{Wannier90} package by projecting the DFT wavefunctions to the $d$ orbital of Pd(Pt) atoms. We then developed a novel method to get the BdG Hamiltonian by adding the pairing term directly to the Wannier tight binding model under the band basis. For the sake of simplicity, we set the pairing term in the same form for all bands, however, our method could be used in more generalized pairing configurations. With such obtained Wannier tight-binding and BdG Hamiltonians, we were then able to calculate surface spectral functions and topological invariants with the \textit{WannierTools} package \cite{Wannier}.

\section{Pairing symmetry classification}\label{APb}

  To find the possible pairing states and analyze their quasi-particle spectrum, one first has to classify all the pairing channels according to the point group symmetry. For multi-dimensional representations, symmetry has to be further identified by finding the local minima (stationary points) of the corresponding Ginzburg-Landau free energy. Resulting symmetry is described by the isotropy group, which can then be used to find the possible quasi-particle gap structures \cite{RevModPhy,Venderbos,volovik}. For the normal state in the presence of SOC, the full symmetry group should be $G_{total}=G\times U(1)\times \mathcal{T}$, where $G$ is the double space group of the lattice. In the vicinity of $T_c$, the symmetry group is spontaneously lowered into the subgroup $G'_{total}$ of $G_{total}$. Explicitly speaking, $g\in G'_{total}$ should be an element of $H\subset G$, accompanied by a phase factor from $U(1)$ or a complex conjugate from $\mathcal{T}$. We list these in table~\ref{tb3} in the form of isotropy group $H(\Gamma_i)$. It can be understood that if a pairing is invariant under $H(\Gamma_i)$, it transforms as if it is the $i$-th irreducible representation (IR) of $H$.

  The $H$ reflects the symmetry of energetics $E_{\bm{k}}$, which can be probed in spectroscopic measurements. We did not consider any further mixing due to the symmetry breaking of the pairing potential \cite{RevModPhy}. 

  \textit{$A_1$ pairing---}    This pairing can be connected to the trivial s-wave pairing. Nodal structure is accidental. In this case, as argued in the main text, it is a DIII class topological superconductor, supporting chiral majorana fermions at its surface.

  \textit{$A_2$ pairing---}    In this phase, $C_{2,xy}$ and $C_{4,z}$ are accompanied by an $e^{i\pi}$ phase. This forces nodes in pairing potential, which is discussed in the next section.

  \textit{$E$ pairing---}  In the $E$-pairing case, the homogeneous GL free energy density can be expressed as
  \begin{equation}
    \begin{aligned}
      f =&\ \alpha|\bm{\eta}|^2 + \beta_1|\bm{\eta}|^4 + \beta_2(\eta_1^*\eta_2-\eta_2^*\eta_1)^2 \\
      & + \gamma_1|\bm{\eta}|^6 + \gamma_2|\bm{\eta}|^2|\eta_1^2+\eta_2^2|^2 + \gamma_3|\eta_1^2||3\eta_2^2-\eta_1^2|^2
    \end{aligned}
  \end{equation}
  where $\bm{\eta}=(\eta_1,\eta_2)$ is the two component order parameter, and ${\alpha,\beta,\gamma}$ are material dependent parameters. The sixth order term has to be included to split accidental degeneracy. Local minima of the free energy can be found at $(1,0)$, $(0,1)$ and $(1,i)$. A multi-component order parameter allows a complex configuration $(1,i)$, spontaneously breaking TR symmetry. 

  \textit{$T_1/T_2$ pairing. ---} In both cases, the order parameter has three components $\bm{\eta}=(\eta_1,\eta_2,\eta_3)$, and the GL free energy density can be written as $f = \alpha|\bm{\eta}|^2 + \beta_1|\bm{\eta}|^4 + \beta_2 |\bm{\eta}\cdot\bm{\eta}|^2 + \beta_3(|\eta_1|^2|\eta_2|^2+|\eta_2|^2|\eta_3|^2+|\eta_3|^2|\eta_1|^2) $. Now the distinct ground states are: $(1,0,0)$, $(1,1,1)$, $(1,\omega,\omega^2)$ and $(1,i,0)$, with $\omega=e^{i2\pi/3}$. Again, the last two terms are related to the time-reversal symmetry breaking (TRSB) phase. According to Table~\ref{tb3}, the saddle points of multi-dimensional IRs are either nematic or TRSB, and only A$_{1/2}$ preserves the full point group symmetry and TR.

  \begin{table*}
      \centering
      \resizebox{\textwidth}{!}{\begin{tabular}{cccccc}
      \hline \hline
          IR  &  Stationary points &  Time-Reversal & Isotropy group $H$ & Node type & $\Delta_{\bm{k}}$  \\
          \hline
           $A_1$  &   --  & TR  &    $O(A_1):\{C_{3,xyz},C_{4,z},C_{2,xy}\}$        & -- & 1, $k_x^2+k_y^2+k_z^2$ \\ \hline
           $A_2$  &   --  &  TR &    $O(A_2)\{C_{3,xyz},-C_{4,z},-C_{2,xy}\}$        & $L$ & $(k_x^2-k_y^2)(k_y^2-k_z^2)(k_z^2-k_x^2)$ \\ \hline
              $E$           &   $(1,0)$  &  TR  &    $D_4(B_1):\{-C_{4,z},-C_{2,xy}\}$        & $L$  &  $\sqrt{3}(k_x^2-k_y^2)$\\
           \multirow{2}{*}{}&   $(0,1)$  &  TR  &    $D_4(A_1):\{C_{4,z},C_{2,xy}\}$        & --  &   $2k_z^2-k_x^2-k_y^2$\\
                            &   $(1,i)$  & TRSB &    $O(E):\{\omega C_{3,xyz},\omega^{2}C_{4,z}\mathcal{K},\omega^{2}C_{2,xy}\mathcal{K}\}$            &  $P$  &  $ k_x^2 + \omega k_y^2 + \omega^2 k_z^2$ \\ \hline
           $T_1$         &   $(1,0,0)$  &  TR  &     $D_4(A_2):\{C_{4,x},-C_{2,yz}\}$           &  $L$   &  $k_yk_z(k_y^2-k_z^2)$\\
            \multirow{3}{*}{}&   $(1,1,1)$  & TR &     $D_3(A_2):\{C_{3,xyz},-C_{2,x\bar{y}}\}$         & $L$ & $k_yk_z(k_y^2-k_z^2)+k_zk_x(k_z^2-k_x^2)+k_xk_y(k_x^2-k_y^2)$\\
                             &   $(1,\omega,\omega^2)$  & TRSB  &    $D_3(E):\{\omega C_{3,xyz},-\omega C_{2,x\bar{y}}\mathcal{K}\}$     & $P$ &  $k_yk_z(k_y^2-k_z^2)+\omega k_zk_x(k_z^2-k_x^2)+\omega^2k_xk_y(k_x^2-k_y^2)$\\
                             &   $(1,i,0)$  & TRSB &     $D_4(E):\{iC_{4,z},-C_{2,y}\mathcal{K}\}$         & $L$ &  $k_yk_z(k_y^2-k_z^2)+ik_zk_x(k_z^2-k_x^2)$\\ \hline
           $T_2$         &   $(1,0,0)$  &  TR  &     $D_4(B_2):\{-C_{4,x},-C_{2,y}\}$           &  $L$  & $k_yk_z$\\
            \multirow{3}{*}{}&   $(1,1,1)$  & TR &     $D_3(A_1):\{C_{3,xyz},C_{2,x\bar{y}}\}$         &  -- &   $k_yk_z+k_zk_x+k_xk_y$\\
                             &   $(1,\omega,\omega^2)$  & TRSB &     $D_3(E):\{\omega C_{3,xyz},\omega C_{2,x\bar{y}}\mathcal{K}\}$     &  $P$ & $k_yk_z+\omega k_zk_x+\omega^2k_xk_y$\\
                             &   $(1,i,0)$  & TRSB &       $D_4(E):\{iC_{4,z},-C_{2,y}\mathcal{K}\}$               &  $L$ &   $k_yk_z+ik_zk_x$\\            
          \hline \hline
      \end{tabular}}
      \caption{The local minima of various pairing states. TR and TRSB indicate time-reversal and time-reversal symmetry breaking phases, respectively. The isotropy group $H$ describes the symmetry of the minima, which is isomorphic to the subgroup of the $O$. The explicit form of the generators is also given. For the node type in single band pairing, `$P$' denotes symmetry protected point nodes, `$L$' denotes line nodes, and `--' denotes that there is possibility of finding accidental nodes. Examples of basis functions are given in the last column. These may have accidental nodes that will be eliminated when higher-order terms of $\bm{k}$ are included. Also note that when projected to a specific band, the pairing function should always be an even function of $\bm{k}$. Here $\omega=e^{i2\pi/3}$. }\label{tb3}
  \end{table*}

\section{Symmetry of the pairing function}\label{APc}
    Time-reversal symmetry can be written as $\mathcal{T}=C\mathcal{K}$ with $\mathcal{T}^2=-1$ (spinful) so that
        \begin{equation}\label{aeq021}
            C_{\bm{k}}C_{\bm{k}}^\dagger=1\quad C_{\bm{k}}=-C_{-\bm{k}}^\text{T}, \quad \mathcal{K}a=a^*\mathcal{K}, \quad  \text{for}\;  a\in\mathbb{C}.
        \end{equation}
    Then we have:
        \begin{equation}\label{aeq03}
          \mathcal{T}H\mathcal{T}^{-1}=H \quad\Rightarrow\quad C_{-\bm{k}}H_{-\bm{k}}^*C_{-\bm{k}}^\dagger=H_{\bm{k}}.
        \end{equation}
    Also, TR symmetry has the properties that given a space group symmetry $g\in G$:
        \begin{equation}\label{sp2}
          g\mathcal{T}\ket{\psi_{\bm{p},\alpha}}=\mathcal{T}g\ket{\psi_{\bm{p},\alpha}}\Rightarrow CD(g)^*=D(g)C.
        \end{equation}
    This indicates that the $C$ matrix can change a representation to its conjugate (adjoint) partner. 

    Now given a superconducting BdG Hamiltonian:
      \begin{equation}
         \mathcal{H}_{\text{BdG}}\qty(\bm{k})=\mqty(H_{\bm{k}}&\Delta_{\bm{k}}\\\Delta_{\bm{k}}^\dagger&-H^T_{-\bm{k}}),
      \end{equation}
    it is very helpful to define a TR basis when TR symmetry is present:
    \begin{equation}\label{sp5}
      \Psi_{\bm{k}}=\qty[\psi_{\bm{k}},\mathcal{T}\psi_{\bm{k}}^\dagger\mathcal{T}^{-1}]^T=\qty[\psi_{\bm{k}},C_{\bm{k}}^T\psi_{\bm{-k}}^\dagger]^T.
    \end{equation}
    In this basis, the BdG Hamiltonian now reads:
    \begin{equation}\label{sp6}
        \begin{aligned}
            \mathcal{H}_{\text{BdG}}\qty(\bm{k}) &= \mqty(\mathbb{1}&\\&C_{\bm{k}}^T)\mqty(H_{\bm{k}}&\Delta_{\bm{k}}\\\Delta_{\bm{k}}^\dagger&-H^T_{-\bm{k}})\mqty(\mathbb{1}&\\&C_{\bm{k}}^*) \\
            & = \mqty(H_{\bm{k}}&\tilde{\Delta}_{\bm{k}}\\\tilde{\Delta}_{\bm{k}}&-H_{\bm{k}}) = H_{\bm{k}}\otimes\tau_z + \tilde{\Delta}_{\bm{k}}\otimes\tau_x.
        \end{aligned}.
    \end{equation}
    Note that the last line only holds when TR symmetry is present. Now the new pairing function, $\tilde{\Delta}_{\bm{k}}=\Delta_{\bm{k}} C_{\bm{k}}^*$ has the following properties required by TR/space group symmetry and Fermi statistics:
            \begin{equation}\label{sp4}
                \begin{aligned}
                   &\mathcal{T}:\quad \tilde{\Delta}_{\bm{k}}^\dagger =\tilde{\Delta}_{\bm{k}}, \\
                   &g:\quad D(g)\tilde{\Delta}_{\bm{k}}D(g)^\dagger = \tilde{\Delta}_{g\bm{k}}, \\
                   \text{Fermi }&\text{statistics}:\quad \tilde{\Delta}_{\bm{k}} = C_{-\bm{k}}\tilde{\Delta}^T_{-\bm{k}} C_{-\bm{k}}^\dagger
                \end{aligned}
            \end{equation}

    Then we can expand this function $\tilde{\Delta}_{\bm{k}}=\sum_{i,\mu}\eta_{i\mu}\Gamma_{i\mu}(\bm{k})$ into IRs of the point group. $\eta_{i\mu}$ is the corresponding order parameter in Ginzburg-Landau (GL) free energy, which is real or complex when TR symmetry is present or absent, respectively.

    At general points in chiral superconductors, strong asymmetric spin-orbit coupling lifts all degeneracy and superconductivity pairs electrons within a single band. This is also in line with the superconductivity fitness condition of \cite{fitness,Ramires,Frigeri}, which states that pairing is most stable when $H_0\left(\bm{k}\right)$ and $\tilde{\Delta}_0\left(\bm{k}\right)$ are simultaneously diagonalized: $\left[H_0\left(\bm{k}\right), \tilde{\Delta}_0\left(\bm{k}\right)\right]=0$. In this eigen-band basis we can write the simple Bogoliubov-de Gennes Hamiltonian as
      
      \begin{equation}\label{BdG_simple}
          \mathcal{H}_{\text{BdG}}\qty(\bm{k})=\mqty(\text{diag}[\xi_{\nu, \bm{k}}] & \text{diag}[\Delta_{\nu, \bm{k}}] \\ \text{diag}[\Delta^*_{\nu, \bm{k}}] & -\text{diag}[\xi_{\nu, \bm{k}}] ).
      \end{equation}
    Pairing is within a single band in this case, which is a single-valued function of $\bm{k}$. TR matrix $C$ is an odd function of $\bm{k}$, such that $t(\bm{k})=-t(\bm{k})$ and $|t(\bm{k})|=1$. For the pairing function $\tilde{\Delta}(\bm{k})$, it is an even function of $\bm{k}$. We discuss such pairing in the computations within the main text. For such single band pairing, symmetries with an additional $U(1)$ phase will stabilize nodes on the Fermi surface. For example, in an $A_2$ phase, there is a $\pi$ phase accompanied with $C_{4,z}$ rotation. Then, there must be at least two line nodes traversing through the rotation axis, since the pairing potential is a real function of $\bm{k}$. For TRSB phase such as the $(1,i)$ for $E$, the $C_3$ has a $e^{i2\pi/3}$ phase. However, since the time-reversal symmetry is broken, we can only conclude that the pairing function is zero at the intersection between the rotation axis and Fermi surface, as the phase is not well defined only at this point. We list all the node types in Table~\ref{tb3}.

    On the Brillouin zone boundary, non-symmorphic symmetry combined with the time-reversal symmetry will force Kramers-type degeneracy. If a line node as classified in the last paragraph resides in this degenerate point, it is then necessary to ask whether this node is susceptible to inter-band pairing. Luckily, according to the results in Table~\ref{tb3}, the line nodes in general are not located on the Brillouin zone boundary, thus inter-band pairing does not affect our result.

    For the results of the superconducting case in the main text, we first adopted the basis defined in Eq.~\ref{sp6} to write the BdG Hamiltonian: $\mathcal{H}_{\bm{k}}=H_{\bm{k}}\otimes\tau_z + \tilde{\Delta}_{\bm{k}}\otimes\tau_x$. For the A$_1$ case, the pairing takes the form 
    \begin{equation}
    	\begin{aligned}
    		 \tilde{\Delta}_{\bm{k}} =& \big\{d_0 + d_1\big[6-2\cos(k_x)-2\cos(k_y) \\
    		       &-2\cos(k_z)\big]\big\}\mathbb{1}_{8\times8}.
    	\end{aligned}
    \end{equation}
     And the parameters for DIII topological phases $(N=4)$ (Fig.~\ref{Fig3}) are $(d_0, d_1)=(-0.05, 0.005)$ (in units of eV). For A$_2$ pairing, it is taken as 
    \begin{equation}
    	\begin{aligned}
    		\tilde{\Delta}_{\bm{k}} =\ & d_0 (\cos k_xa-\cos k_ya) \\
    		&\times(\cos k_ya-\cos k_za)(\cos k_za-\cos k_xa)\mathbb{1}_{8\times8},
    	\end{aligned}
    \end{equation}
    where the size of $d_0$ doesn't affect the physics and is set to be 0.02 eV.

\section{Tight-binding Hamiltonian}\label{APd}
  \subsection{Normal state}
            As is mentioned in the main text, we constructed a tight binding model from the $\bar{E}_1$ representation at Wyckoff position $4a$ to give a minimal model that is compatible with space group symmetry. This can be understood as a model describing s-wave-like spinful orbitals originating from distorted B(Pd,Pt)$_6$ octahedra centered at Wyckoff position 4a. This model is simple yet sufficient to capture the topology and symmetry of the band structure. Wyckoff position 4a is a maximal Wyckoff position \cite{bilbao1,bilbao2,bilbao3,bilbao4,TQChem}, and the resulting representation is elementary. By following the spinful basis $\qty(\ket{1},\ket{2},\ket{3},\ket{4})\otimes\qty(\ket{\uparrow},\ket{\downarrow})$, we can find all the allowed hopping terms according to their relative position and check whether they are consistent with space group symmetry.

            The exact form of the tight binding model is shown as follows:
\begin{widetext}
	\begin{equation}\label{aeq1}
	  H = H(\bm{k})=
	  \begin{pmatrix}
	    h_0(\bm{k})-\mu & h_{12}(\bm{k}) & h_{13}(\bm{k}) & h_{14}(\bm{k}) \\
	    h_{12}(\bm{k})^\dagger & h_0(\bm{k})-\mu & h_{23}(\bm{k}) & h_{24}(\bm{k}) \\
	    h_{13}(\bm{k})^\dagger & h_{23}(\bm{k})^\dagger & h_0(\bm{k})-\mu & h_{34}(\bm{k}) \\
	    h_{14}(\bm{k})^\dagger & h_{24}(\bm{k})^\dagger & h_{34}(\bm{k})^\dagger & h_0(\bm{k})-\mu
	  \end{pmatrix},
	\end{equation}
    with

	\begin{equation}
	    h_0(\bm{k})   = -2t_2[\cos(k_x)+\cos(k_y)+\cos(k_z)]\sigma_0,
	\end{equation}
     
	\begin{equation}
		\begin{aligned}
		            h_{12}(\bm{k})  &= [t_0(e^{ik_y}+e^{ik_y+ik_z})+t_1(1+e^{ik_z} + e^{ik_x+ik_y} + e^{ik_x+ik_y+ik_z})]\sigma_0 \\+ &\begin{pmatrix}
	                                                                                   d_2e^{ik_y}+d_2e^{ik_y+ik_z}-d_3 -d_3e^{ik_z}+d_4e^{ik_x+ik_y}+d_4e^{ik_x+ik_y+ik_z}\\
	                                                                                   -d_2e^{ik_y}-d_2e^{ik_y+ik_z}-d_4 -d_4e^{ik_z}+d_3e^{ik_x+ik_y}+d_3e^{ik_x+ik_y+ik_z}\\
	                                                                                   -d_1e^{ik_z}+d_1e^{ik_y+ik_z}-d_5+d_5e^{ik_z}-d_5e^{ik_x+ik_y}+d_5e^{ik_x+ik_y+ik_z}
	                                                                                 \end{pmatrix}\cdot i\va{\sigma},
		\end{aligned}
	\end{equation}
	\begin{equation}
	  \begin{aligned}
		            h_{13}(\bm{k})  &= [t_0(e^{ik_x}+e^{ik_x+ik_y})+t_1(1+e^{ik_y} + e^{ik_z+ik_x} + e^{ik_x+ik_y+ik_z})]\sigma_0 \\+ &\begin{pmatrix}
	                                                                                   -d_2e^{ik_x}-d_2e^{ik_x+ik_y}-d_4 -d_4e^{ik_y}+d_3e^{ik_z+ik_x}+d_3e^{ik_x+ik_y+ik_z}\\
	                                                                                   -d_1e^{ik_x}+d_1e^{ik_x+ik_y} -d_5+d_5e^{ik_y}-d_5e^{ik_z+ik_x}+d_5e^{ik_x+ik_y+ik_z}\\
	                                                                                   d_2e^{ik_x}+d_2e^{ik_x+ik_y}-d_3-d_3e^{ik_y}+d_4e^{ik_z+ik_x}+d_4e^{ik_x+ik_y+ik_z}
	                                                                                 \end{pmatrix}\cdot i\va{\sigma}, 
		\end{aligned}                                                    	
	\end{equation}

	\begin{equation}
		\begin{aligned}
	            h_{14}(\bm{k})  &= [t_0(e^{ik_z}+e^{ik_z+ik_x})+t_1(1+e^{ik_x} + e^{ik_y+ik_z} + e^{ik_x+ik_y+ik_z})]\sigma_0 \\+ &\begin{pmatrix}
	                                                                               -d_1e^{ik_z}+d_1e^{ik_z+ik_x} -d_5+d_5e^{ik_x}-d_5e^{ik_y+ik_z}+d_5e^{ik_x+ik_y+ik_z}\\
	                                                                               d_2e^{ik_z}+d_2e^{ik_z+ik_x} -d_3-d_3e^{ik_x}+d_4e^{ik_y+ik_z}+d_4e^{ik_x+ik_y+ik_z}\\
	                                                                               -d_2e^{ik_z}-d_2e^{ik_z+ik_x}-d_4-d_4e^{ik_x}+d_3e^{ik_y+ik_z}+d_3e^{ik_x+ik_y+ik_z}
	                                                                             \end{pmatrix}\cdot i\va{\sigma}, 
		\end{aligned}
	\end{equation}
	\begin{equation}
		\begin{aligned}
		        h_{23}(\bm{k})  &= [t_0(1+e^{ik_x})+t_1(e^{-ik_y}+e^{-ik_z}+e^{ik_x-ik_z}+e^{ik_x-ik_y})]\sigma_0 \\+ &\begin{pmatrix}
		                                                                           -d_1+d_1e^{ik_x}+d_5e^{ik_z-ik_x}\\
		                                                                           -d_2-d_2e^{ik_x} +d_3e^{ik_z-ik_x}\\
		                                                                           -d_2-d_2e^{ik_x}-d_4e^{ik_z-ik_x}
		                                                                         \end{pmatrix}\cdot i\va{\sigma},
		\end{aligned}
	\end{equation}
	\begin{equation}
		\begin{aligned}
		            h_{24}(\bm{k})  &= [t_0(1+e^{ik_y})+t_1(e^{ik_x}+e^{ik_z}+e^{ik_x-ik_y}+e^{ik_z-ik_y})]\sigma_0 \\+ &\begin{pmatrix}
		                                                                               d_2+d_2e^{ik_y}+d_4e^{ik_x-ik_y}\\
		                                                                               d_1-d_1e^{ik_y} -d_5e^{ik_x-ik_y}\\
		                                                                               d_2+d_2e^{ik_z}-d_3e^{ik_x-ik_y}
		                                                                             \end{pmatrix}\cdot i\va{\sigma},
		\end{aligned}
	\end{equation}
	\begin{equation}
		\begin{aligned}
		            h_{34}(\bm{k})  &= [t_0(1+e^{ik_z})+t_1(e^{-ik_x}+e^{-ik_y}+e^{ik_z-ik_x}+e^{ik_z-ik_y})]\sigma_0 \\+ &\begin{pmatrix}
		                                                                               -d_2-d_2e^{ik_z} +d_3e^{ik_y-ik_z}\\
		                                                                               -d_2-d_2e^{ik_z} -d_4e^{ik_y-ik_z}\\
		                                                                               -d_1+d_1e^{ik_z} +d_5e^{ik_y-ik_z}
		                                                                             \end{pmatrix}\cdot i\va{\sigma}.
		\end{aligned}
	\end{equation}

\end{widetext}
      Here $h_{ij}$ are the hopping matrices acting on atomic sites, and $\sigma_i$ acts on spin basis. The minimal parameter set we adopted here is $(t_0,t_1,t_2,d_1,d_2,d_3,d_4,d_5,\mu)$.  $\mu$ is the chemical potential; $t_i$ is the $i-$th nearest neighbor normal hopping, $d_{1/2}$ is the nearest neighbor SOC, and $d_{3/4/5}$ is the next nearest neighbor SOC.

  \section{k$\cdot$p model and explicit matrix representations of the little groups}\label{APe}

    To analyze the low energy effect model of  multi-fold fermions, it is ideal to construct a $\bm{k}\cdot\bm{p}$ model. We focus on the linear order in $\bm{k}$, so that the Hamiltonian can be written in the form:
      \begin{equation}
        H(\delta\bm{k}) = \sum_{i=1,2,3}M^i\delta\bm{k}_i,
      \end{equation}
    with $\delta\bm{k} = \bm{k}-\bm{K}_{\Gamma/R}$. The induced representation would be $(\bar{E}_1\uparrow G)\downarrow G_\Gamma \approx \bar{\Gamma}_6(2)\oplus\bar{\Gamma}_7(2)\oplus\bar{\Gamma}_8(4)$ at the $\Gamma$ point and $(\bar{E}_1\uparrow G)\downarrow G_R \approx\bar{R}_6(2)\oplus\bar{R}_7\bar{R}_8(6)$ at the $R$ point, which can be found at \textit{Bilbao Crystallographic server} \cite{bilbao4}. The numbers in the parentheses indicate the numbers of degeneracy, and $\bar{R}_7$ and $\bar{R}_8$ are related by TR symmetry. Here, $G_{\Gamma/R}$ denotes the little group of point $\Gamma/R$.  This means that the four-fold fermion is subject to $\bar{\Gamma}_8(4)$, and the two three-fold fermions are subject to $\bar{R}_7(3),\bar{R}_8(3)$, respectively. Then, to find the $\bm{k}\cdot\bm{p}$ Hamiltonian, for example, of the four-fold fermion at $\Gamma$, one just needs to find one set of matrix representations $D(g)$ of $\bar{\Gamma}_8(4)$ for $g\in G_{\Gamma}$ explicitly and then construct a 4-band $\bm{k}\cdot\bm{p}$ Hamiltonian $H_{\bm{k}}$ that is invariant, such that $D(g)H_{g^{-1}\bm{k}}D(g)^\dagger = H_{\bm{k}}$. Then, all the allowed terms can be derived from this relation.

  \subsection{Four-fold fermion at $\Gamma$}
    For $\bar{\Gamma}_8$ of $G_{\Gamma}$, we use the following form \cite{bilbao4}:
 \begin{widetext}
    \begin{equation}\label{aeqIR1}
      \begin{aligned}
       & \qty{1|0}=\mathbb{1}_{4\times4},   \qquad\qty{2_{001}|\tfrac{1}{2}0\tfrac{1}{2}}=\qty(\smqty{-i&&&\\&i&&\\&&i&\\&&&-i}),   \qquad\qty{2_{010}|0\tfrac{1}{2}\tfrac{1}{2}}= \qty(\smqty{&-1&&\\1&&&\\&&&-i\\&&-i&}), \\  & \qty{1^\text{d}|0}=-\mathbb{1}_{4\times4},  \qquad \qty{3^+_{111}|0}=\frac{\sqrt{2}}{2}\qty(\smqty{\e^{i5\pi/12}&\e^{-i\pi/12}&&\\\e^{i5\pi/12}&\e^{i11\pi/12}&&\\&&\e^{-i5\pi/12}&\e^{i7\pi/12}\\&&\e^{-i11\pi/12}&\e^{-i11\pi/12}}), \qquad\qty{2_{110}|\tfrac{1}{4}\tfrac{3}{4}\tfrac{1}{4}}=\qty(\smqty{&&-1&\\&&&-1\\1&&&\\&1&&}).
      \end{aligned}
    \end{equation}

    Here $\qty{1^\text{d}|0}$ denotes the $2\pi$ rotation. We first get two sets of matrices so that $M^i = v_1F_1^i + v_2F_2^i$:
        \begin{equation}\label{eeq4}
          \begin{aligned}
                F_1^1 & = \mqty(\dmat{\sigma_x,\sigma_y}),\ F_1^2=\mqty(\dmat{\sigma_y,\sigma_x}),\ F_1^3=\mqty(\dmat{\sigma_z,-\sigma_z}),   \\
                F_2^1 & = \qty(\smqty{0&0&e^{i\pi/12}&0 \\ 0&0&0&-ie^{i\pi/12} \\e^{-i\pi/12}&0&0&0 \\0&ie^{-i\pi/12}&0&0}), \ F_2^2=\qty(\smqty{0&0&-e^{-i\pi/12}&0\\0&0&0&-ie^{-i\pi/12}\\-e^{i\pi/12}&0&0&0\\0&ie^{-i\pi/12}&0&00}),\\
                  &F_2^3=\qty(\smqty{0&0&0& e^{i\pi/4}\\0&0& e^{-i\pi/4}&0\\0& e^{i\pi/4}&0&0\\ e^{-i\pi/4}&0&0&0 }).
          \end{aligned}
        \end{equation}
\end{widetext}

    The four-fold fermion at the $\Gamma$ point of a cubic lattice was previously identified as a spin-3/2 fermion \cite{Bernevig1,Shoucheng,Hassan2,Kim,Brydon,Venderbos}.   To better connect this effective Hamiltonian to the spin-3/2 fermion, we recombine the two sets of $F$ matrices to find the required form: $J^i = -\tfrac{1}{2}F_1^i+F_2^i$, and $T^i = F_1^i+\tfrac{1}{2}F_2^i$, so that $H=\bm{k}\cdot(v'_1\vb{J}+v'_2\vb{T})$. In this way, the $J^i$ matrices form a 4-dimensional, spin-3/2 representation of $SO(3)$: $\qty[J^i,J^j]=i\epsilon^{ijk}J^k$ and $\sum J^2 = \sqrt{\tfrac{1}{2}l(l+1)}\mathbb{1}$ with $l=3/2$. And the remaining $T$ matrices introduce the effect of a cubic crystalline field. 

  \subsection{Three-fold fermion at R}

    At the $R$ point, the 6-fold degeneracy is composed of an $\bar{R}_7\bar{R}_8$ representation, where $\bar{R}_7$ and $\bar{R}_8$ are connected by TR symmetry. They are interpreted as two spin-1 fermions. We can first write one of the spin-1 fermion as $H_{\bm{k}}=\bm{k}\cdot\Gamma$. We project the $\Gamma$ matrices to the $\bar{R}_7$ representation, whose explicit form is
    \begin{widetext}
        \begin{equation}\label{aeqIR2}
          \begin{aligned}
           & \qty{1|0}=\mathbb{1}_{3\times3},   \qquad\qty{2_{001}|\tfrac{1}{2}0\tfrac{1}{2}}=\qty(\smqty{1&&\\&1&\\&&-1}),   \qquad\qty{2_{010}|0\tfrac{1}{2}\tfrac{1}{2}}= \qty(\smqty{1&&\\&-1&\\&&1}), \\  & \qty{1^\text{d}|0}=-\mathbb{1}_{3\times3},  \qquad \qty{3^+_{111}|0}=\qty(\smqty{&&-1\\1&&\\&1&}), \qquad\qty{2_{110}|\tfrac{1}{4}\tfrac{3}{4}\tfrac{1}{4}}=\qty(\smqty{&-i&\\-i&&\\&&-i}).
          \end{aligned}
        \end{equation}
    \end{widetext}

    The same can be done for $\bar{R}_8$, whose matrices are the complex conjugate of those of $\bar{R}_7$ since $\bar{R}_8=\bar{R}^*_7$.  The full Hamiltonian can be written as:
        \begin{equation}\label{eeq3.5}
          H_{\delta\bm{k}}=\mqty(a\delta\bm{k}\cdot \Gamma  & be^{i\theta}\delta\bm{k}\cdot \Omega \\be^{-i\theta}\delta\bm{k}\cdot \Omega & a\delta\bm{k}\cdot \Gamma ),
        \end{equation}
    with $a,b,\theta$ being real parameters. The resulting $\Gamma^i$ and $\Omega^i$ matrices are:
        \begin{equation}\label{eeq3}
        \begin{aligned}
          \Gamma^1 &= \mqty(0&0&0\\0&0&-i\\0&i&0), \Gamma^2=\mqty(0&0&-i\\0&0&0\\i&0&0), \Gamma^3=\mqty(0&-i&0\\i&0&0\\0&0&0), \\
          \Omega^1 &= \mqty(0&0&0\\0&0&1\\0&1&0), \Omega^2=\mqty(0&0&-1\\0&0&0\\-1&0&0), \Omega^3=\mqty(0&1&0\\1&0&0\\0&0&0).
          \end{aligned}
        \end{equation}
    TR operator here reads $\mathcal{T}=\smqty(0&\mathbb{1}\\-\mathbb{1}&0)\mathcal{K}$, with $\mathcal{K}$ the conjugate operator. Here, we still have $\mathcal{T}^2=-1$, despite  it being a `spin-1' fermion, as the fundamental block is a spin-1/2 electron. Again, to make the feature of the spin-1 fermion more explicit, we have
        \begin{equation}\label{dec2}
          J^i = \mqty(\dmat{\Gamma^i,\Gamma^{i}}),\qquad T^i=\mqty(\admat{e^{i\theta}\Omega^i,e^{-i\theta}\Omega^i}).
        \end{equation}
    Here $\sum J^2=4\mathbb{1}$, as it represents two copies of a spin-1 fermion. These double spin-1 fermions are also found in previous literature about chiral cubic crystals \cite{Bernevig1,Shoucheng,Hassan2}.

\section{Identification of unconventional fermions}\label{APf}
      The origin of these various types of fermions in lattice systems is from particle physics. Given a relativistic fermion with spin-N, we have a Hamiltonian $H_{\bm{k}}=v(\bm{k}\cdot \bm{J})_{(2N+1)\times(2N+1)}$, with a real coefficient $v$, and $\bm{J} = \{J_x,J_y,J_z\}$ is the (2N+1) dimensional (or spin-N) IR of rotation group $SO(3)$. The energies of the bands are $E_{\bm{k}}=vnk$, with $n=-N,-N+1,...N-1,N$. For the $n$-th band $\ket{n_{\bm{k}}}$, the berry curvature can be calculated as
            \begin{equation}\label{BCnOM1}
                \begin{aligned}
                    \mathcal{F}^n(\bm{k}) &= -i\epsilon^{ijk}\sum_{m\neq n}\qty\Big[\langle\partial_j n_{\bm{k}}{\dyad{m}}\partial_j n_{\bm{k}}\rangle ] \\
                            & =  -i\epsilon^{ijk}\sum_{m\neq n}\qty\Big[\frac{\mel{n}{\partial_iH}{m}\mel{m}{\partial_jH}{n}}{(E_n-E_m)^2}],
                \end{aligned}
            \end{equation}
        whose result can be found by
            \begin{equation}\label{BCnOM2}
              \mathcal{F}^n(\bm{k}) = n\frac{\hat{k}}{k^2}.
            \end{equation}
      Then, after integration over the enclosing surface, the Chern number of the $n$-th band can be found to be $C_{n}=2n$, and the monopole charge can be identified as
        \begin{equation}\label{monopole}
        	\begin{aligned}
        	           C_{\text{monopole}}\  &= \sum_{E_n>0}C_{n}\\
        	           &=\begin{cases}
                                                                \text{sgn}(v)(N^2+N) \qquad &(\text{integer spin})\\
                                                                \text{sgn}(v)(N+1/2)^2 \qquad &(\text{half integer spin})
                                                          \end{cases}
        	\end{aligned}
         \end{equation}

      Due to the crystalline symmetry in these solids, the dispersions of these fermions are distorted. However, the Chern numbers of a given band and degeneracy at high symmetry points are unaffected and can be used to identify such fermions. For example, for the 4-fold fermion No.~14 of Table~\ref{ChiralFermions} in the main text, we numerically computed the Chern numbers of the four bands at a tiny sphere around $\Gamma$, and we found them to be $3,1,-1$ and $-3$ (from lower to the higher band). This indicates that the fermion has spin-3/2 with monopole charge $-4$. Numerical computation was done with the \textit{WannierTools} package \cite{Wannier}.

  \section{Cubic dispersion Majorana bands on the (111) surface of DIII topological superconducting Li$_2$Pd$_3$B}\label{APg}


    In the main text, we have shown that for the (111) surface of DIII topological superconducting Li$_2$PdB$_3$, there is a linear as well as a cubic Majorana cone on each side, manifesting the topological effects of unconventional fermions on the superconducting states. To further study the dispersion relation around the surface $\bar{\Gamma}$, we obtained the eight eigen-states nearest to the zero energy at $\bar{\Gamma}$ and projected the complicated slab Wannier Hamiltonian to them to get the 8-by-8 surface Hamiltonian. The energy spectrum of this projected surface Hamiltonian is shown in Fig.~\ref{Fig3}b, and the linear as well as the cubic dispersion were verified by suitable fitting. It should be noted that the band difference between the two surfaces is due to the lack of inversion symmetry.
    
    It is pointed out in \cite{Fang2015} that the cubic Majorana cone can result from spin-3/2 fermions and is protected by time-reversal $\hat{\mathcal{T}}$, particle-hole $\hat{\mathcal{P}}$ and $\hat{C}_3$ rotation symmetry, which is exactly the case for the (111) surface of Li$_2$Pd$_3$B. The argument is as follows. Assuming that the Majorana surface states are formed from spin-3/2 fermions, the symmetry operators mentioned above and their restrictions on the surface Hamiltonian can be written as
    \begin{equation}
        \begin{aligned}
      \hat{\mathcal{T}} = i\sigma_{y}\hat{\mathcal{K}},\ \ \ \ & \hat{\mathcal{T}}^{-1}H(\bm{k})\hat{\mathcal{T}} = H(-\bm{k}),\\
      \hat{\mathcal{P}} = \sigma_{x}\hat{\mathcal{K}},\ \ \ \ & \hat{\mathcal{P}}^{-1}H(\bm{k})\hat{\mathcal{P}} = -H(-\bm{k}),\\
      \hat{C}_{3} = -\sigma_{0},\ \ \ \ & \hat{C}_{3}^{-1}H(\hat{C}_{3}\bm{k})\hat{C}_{3} = H(\bm{k}).
    \end{aligned}
    \end{equation}

    Simple calculation tells us that $\hat{\mathcal{T}}$ together with $\hat{\mathcal{P}}$ requires that the surface Hamiltonian should only contain $k^{(odd)}\sigma_{x,y}$ terms while $\hat{C}_{3}$ requires that $H(k_+, k_-)=H(k_+e^{i\frac{2}{3}\pi}, k_-e^{-i\frac{2}{3}\pi})$ with $k_{\pm}=k_x\pm ik_y$. Thus, to the lowest-allowed $k$ terms, the surface Majorana Hamiltonian is 
    \begin{equation}
      H(k_+, k_-) = c_1(k_{+}^3\sigma_+ + k_{-}^3\sigma_-) + c_2(k_{+}^3\sigma_- + k_{-}^3\sigma_+),
    \end{equation}
    where $\sigma_{\pm} = \sigma_x \pm i\sigma_y$. In the contrast, for spin-1/2 fermions, the lowest-allowed terms are linear in $k$. In the weak pairing limit, the surface Majorana states are just linear combinations of particle and hole states near the Fermi level of bulk Li$_2$Pd$_3$B. In this sense, the existence of cubic Majorana bands is the direct result of the spin-3/2 unconventional fermions on the Fermi surfaces of Li$_2$Pd$_3$B.


\begin{thebibliography}{99}
\bibitem{Yulin} N. B. M. Schroter et al., Nat. Phys. {\bf 15}, 759-765 (2019).

\bibitem{Hassan3} D. S. Sanchez et al., Nature {\bf 567}, 500-505 (2019).

\bibitem{HongDing} Z. Rao et al., Nature {\bf 567}, 496-499 (2019).

\bibitem{Takafumi} D. Takane et al., Phys. Rev. Lett. {\bf 122}, 076402 (2019).

\bibitem{Gorkov1}
L. P. Gor'kov, E. I. Rashba, Phys. Rev. Lett. {\bf87}, 037004 (2001).

\bibitem{Sigrist1}
P. A. Frigeri, D. F. Agterberg, A. Koga, and M. Sigrist, Phys. Rev. Lett. {\bf92}, 097001 (2004).

\bibitem{Sigrist3}
P. A. Frigeri, D. F. Agterberg, and M. Sigrist, New J. Phys. {\bf6}, 115 (2004).

\bibitem{Agterberg2003}
D. F. Agterberg, Physica C {\bf 387}, 13–6 (2003).

\bibitem{Sigrist2005}
R. P. Kaur, D. F. Agterberg and M. Sigrist, Phys. Rev. Lett. {\bf94}, 137002 (2005).

\bibitem{Feigelman2003}
O. V. Dimitrova and M. V. Feigel’man, JETP Lett. {\bf78}, 637 (2003).



\bibitem{Mukuda}
H. Mukuda, et al., Phys. Rev. Lett {\bf104}, 017002 (2010).

\bibitem{Yogi}
Yogi, Mamoru, et al., J. Phys. Soc. J. {\bf75}, 013709 (2006).

\bibitem{Kimura}
Kimura, Noriaki, et al., Phys. Rev. Lett {\bf98}, 197001 (2007).

\bibitem{Lu2015}
J. M. Lu et al., Science {\bf350}, 1353-1357 (2015).


\bibitem{Xi2016}
X. Xi et al., Nat. Phys. {\bf12}, 139-143 (2016).

\bibitem{Nagaosa2012}
S. Nakosai, Y. Tanaka and N. Nagaosa, Phys. Rev. Lett. {\bf108}, 147003 (2012).

\bibitem{Schmalian2015}
M. S. Scheurer and J. Schmalian, Nat. Commun. {\bf6}, 6005 (2015).

\bibitem{Fujimoto2010}
M. Sato and S. Fujimoto,
Phys. Rev. Lett. {\bf105}, 217001 (2010).



\bibitem{Samokhin2015}
K. V. Samokhin, 
Ann. Phys. {\bf359}, 385 (2015).

\bibitem{Zhang2011}
X. L. Qi and S. C. Zhang, 
Rev. Mod. Phys. {\bf83}, 1057 (2011).




\bibitem{Smidman} M. Smidman, M. B. Salamon, H. Q. Yuan, and D. F. Agterberg, Rep. Prog. Phys. {\bf 80}, 036501 (2017)

\bibitem{Hassan1} G. Chang et al., Nat. Mater. {\bf 17}, 978 (2018).

\bibitem{Bernevig1} B. Bradlyn et al., Science {\bf 353}, 6299 (2016).

\bibitem{Shoucheng} P. Tang, Q. Zhou, and S. C. Zhang, Phys. Rev. Lett. {\bf 119}, 206402 (2017).


\bibitem{Hassan2} G. Chang et al., Phys. Rev. Lett. {\bf 119}, 206401 (2017).




\bibitem{Hirata} K. Togano, P. Badica, Y. Nakamori, S. Orimo, H. Takeya, and K. Hirata, Phys. Rev. Lett. {\bf 93}, 247004 (2004).

\bibitem{Yuan} H. Q. Yuan, D. F. Agterberg, N. Hayashi, P. Badica, D. Vandervelde, K. Togano, M. Sigrist, and M. B. Salamon, Phys. Rev. Lett. {\bf 97}, 017006 (2006).


\bibitem{ZhengGuoqing1} M. Nishiyama, Y. Inada, and G. Q. Zheng, Phys. Rev. B {\bf71}, 220505(R) (2005).

\bibitem{ZhengGuoqing2} M. Nishiyama, Y. Inada, and G. Q. Zheng, Phys. Rev. Lett. {\bf 98}, 047002 (2007).

\bibitem{ZhengGuoqing3} S. Harada, J. J. Zhou, Y. G. Yao, Y. Inada, and G. Zheng, Phys. Rev. B {\bf 86}, 220502(R) (2012).


\bibitem{SpecificHeat} H. Takeya, K. Hirata, K. Yamaura, K. Togano, M. El Massalami, R. Rapp, F. A. Chaves, and B. Ouladdiaf, Phys. Rev. B {\bf 72}, 104506 (2005).

\bibitem{SpecificHeat2} G. Eguchi, D. C. Peets, M. Kriener, S. Yonezawa, G. Bao, S. Harada, Y. Inada, G. Q. Zheng, and Y. Maeno, Phys. Rev. B {\bf87}, 161203(R) (2013).


\bibitem{Pickett} K.-W. Lee, and W. E. Pickett, Phys. Rev. B {\bf 72}, 174505 (2005).



\bibitem{Schnyder2} A. P. Schnyder and P. M. R. Brydon, J. Phys. Condens. Matter {\bf27}, 243201 (2015).

\bibitem{Ryu} A. P. Schnyder, and S. Ryu, Phys. Rev. B {\bf 84}, 060504(R) (2011).

\bibitem{Schnyder2012}
A. P. Schnyder, P. M. R. Brydon, and C. Timm. 
Phys. Rev. B {\bf85},024522 (2012).




\bibitem{Bradley} C. J. Bradley, B. L. Davies, Rev. Mod. Phys. {\bf40}, 359 (1968).

\bibitem{Sigrist2} A. Ramires, and M. Sigrist, Phys. Rev. B {\bf 94}, 104501 (2016).


\bibitem{RevModPhy} M. Sigrist and K. Ueda, Rev. Mod. Phys. {\bf63}, 239 (1991).

\bibitem{volovik} G. E. Volovik and L. P. Gor'kov, Zh. Eksp. Teor. Fiz, {\bf 88}: 1412-1428, (1985).

\bibitem{Venderbos} J. W. F. Venderbos, L. Savary, J. Ruhman, P. A. Lee, and L. Fu, Phys. Rev. X {\bf8}, 11029 (2018).

\bibitem{Brydon} P. M. R. Brydon, L. Wang, M. Weinert, D. F. Agterberg, Phys. Rev. Lett., {\bf116}(17), (2016).

\bibitem{Mostofi2014} A. A. Mostofi et al., Comput. Phys. Commun. {\bf185}, 2309 (2014)

\bibitem{Ludwig2008}
A. P. Schnyder, S. Ryu, A. Furusaki and A.W.W. Ludwig, 
Phys. Rev. B {\bf78}, 195125 (2008).

\bibitem{Xiaoliang} X.-L. Qi, T. L. Hughes, and S.-C. Zhang, Phys. Rev. B {\bf 81}, 134508 (2010).


\bibitem{Fang2015} C. Fang, B. A. Bernevig and M. J. Gilbert, Phys. Rev. B {\bf 91}, 165421 (2015).

\bibitem{Yang2016} W. Yang, Y. Li, and C. Wu, Phys. Rev. Lett. {\bf117}, 075301 (2016).

\bibitem{Fu2010} L. Fu and E. Berg, Phys. Rev. Lett. {\bf105}, 097001 (2010).
        %

\bibitem{Sato2010} M. Sato, Phys. Rev. B {\bf81}, 220504(R) (2010).

\bibitem{Karki2010} A. B. Karki, et al. Phys. Rev. B {\bf82}, 064512 (2010).

\bibitem{Joshi2015} B. Joshi, et al. J. Phys.: Conf. Ser. {\bf592}, 012069 (2015).

\bibitem{Krupka1969} M. C. Krupka, M. C., et al. J. Less Common Met. {\bf17}, 91-98 (1969).

\bibitem{Simon2004} A. Simon, T. Gulden and Z. Anorg, Allg. Chem. {\bf630}, 2191–2198 (2004).



\bibitem{dft1} P. Hohenberg, W. Kohn, Phys. Rev. B {\bf136} (3B), B864-B871 (1964).
\bibitem{dft2} G. Kresse, J. Furthmuller, Comput. Mater. Sci. {\bf6}, 15-50 (1996).
\bibitem{dft3} P. E. Bl\"ochl,  Phys. Rev. B {\bf50}, 17953-17979 (1994).
\bibitem{dft4} J. P. Perdew, K. Burke, M. Ernzerhof,  Phys. Rev. Lett. {\bf77}, 3865-3868 (1996).
\bibitem{dft5} D. C. Langreth, M. J. Mehl,  Phys. Rev. B {\bf28}, 1809-1834 (1983).

\bibitem{Wannier} Q. Wu, S. Zhang, H. Song, M. Troyer and Alexey A. Soluyanov, Comp. Phys. Comm. {\bf224}, 405-416 (2018).

\bibitem{fitness}A. Ramires and M. Sigrist, Phys. Rev. B {\bf 94}, 104501 (2016).
\bibitem{Ramires}A. Ramires, D. F. Agterberg and M. Sigrist, Phys. Rev. B, {\bf 98}, 024501 (2018).
\bibitem{Frigeri}P. A. Frigeri, et al., Phys. Rev. lett. {\bf 92}, 097001 (2004).

\bibitem{bilbao1} M. I. Aroyo et al., Bulg. Chem. Commun. {\bf 43}(2) 183-197 (2011).
\bibitem{bilbao2} M. I. Aroyo et al., Z. Krist. {\bf 221}, 1, 15-27 (2006). 
\bibitem{bilbao3} M. I. Aroyo et al., {\bf A62}, 115-128 (2006).
\bibitem{bilbao4} L. Elcoro et al., J. of Appl. Cryst. {\bf 50}, 1457-1477 (2017).
\bibitem{TQChem} B. Bradlyn et al., Nature {\bf 547}, 298-305 (2017).

\bibitem{Kim} H. Kim et al., Sci. Adv. {\bf4}, eaao4513 (2018).

\end{thebibliography}

\end{document}